\documentstyle[aps,prl,multicol,epsf]{revtex}
\draft
\begin{document}
\title{Crossover from critical orthogonal to critical unitary statistics
at the Anderson transition}
\author{$^{1,2}$M. Batsch, $^1$L. Schweitzer, $^2$I. Kh. Zharekeshev,
and $^2$B. Kramer}
\address{$^1$Physikalisch-Technische Bundesanstalt, Bundesallee 100,
38116 Braunschweig, Germany\\
$^2$I. Institut f\"ur Theoretische Physik, Universit\"at Hamburg,
Jungiusstra{\ss}e 9, 20355 Hamburg, Germany}
\date{Received 16 April 1996}
\maketitle
\begin{multicols}{2}[%
\begin{abstract}
We report a novel scale-independent, Aharonov-Bohm flux controlled
crossover from critical orthogonal to critical unitary statistics
at the disorder induced metal insulator transition.
Our numerical investigations show that at the critical point the level 
statistics are definitely distinct and determined by fundamental symmetries.
The latter is similar to the behavior of the metallic phase
known from random matrix theory.  The Aharonov-Bohm flux dependent crossover
is characteristic of the critical ensemble.
\end{abstract}
]

In analogy with classical phase transitions it is commonly believed that the
critical behavior at the disorder induced metal-insulator transition (MIT)
is determined by fundamental symmetries.
This is not only the case for the critical exponent
of the localization length and the dc-conductivity, but also for the
statistical properties of energy eigenvalues and the wave functions.
It was in particular the discovery of a new critical universality
class of spectral correlations \cite{Sea93,KLAA94} that challenged our
current understanding of the localization-delocalization transition.

Random matrix theory (RMT) \cite{Meh91} provides a general
statistical description of the eigenvalue fluctuations and
correlations in the metallic phase \cite{Efe83}.
The corresponding universal energy level statistics only depend on fundamental
symmetries of the system under consideration. For time reversal
symmetry  orthogonal (GOE) and symplectic (GSE) Gaussian ensembles are
appropriate, whereas in case of broken time reversal symmetry the unitary
(GUE) Gaussian ensemble describes the level statistics.
In contrast to the metallic phase, the uncorrelated energy levels of
the localized states in the insulating regime are characterized by the
Poisson statistics, independent of the symmetry.

A critical statistics was not only found at the MIT in 3d
in the presence of time reversal symmetry \cite{Sea93,ZK95a,BM95},
but also for 2d symplectic systems \cite{SZ95,Eva95},
where spin rotational invariance is broken but time reversal symmetry is
conserved.
In both cases the critical level statistics for small energy level
separations closely resemble the distributions corresponding to
the metallic phases. The probability density distribution $P_c(s)$ of the
normalized energy separation, $s=|E_n-E_{n+1}|/\Delta$, is still
$P_c(s)\sim s^{\beta}$ for small $s$,
where $\beta=1$ or 4 in the orthogonal and symplectic case, respectively.
Here, $E_n$ and $E_{n+1}$ are two successive eigenenergies and $\Delta$ is
the mean level spacing. 
This reflects the energy level repulsion due to the strong overlap of the 
corresponding eigenstates.
From these results at the critical point and the universal behavior in
the metallic phase for all possible symmetries, one expects that
$P_c(s)\sim s^{\beta}$ with $\beta=2$ for the critical
unitary case, when time reversal symmetry is broken.

Recently, however, it has been claimed \cite{HoS94b} that
the critical level statistics are independent of the presence or
absence of time reversal symmetry. The authors based their assertion
on numerical calculations of the nearest-neighbor energy spacing distribution 
$P_c(s)$ and the $\Delta_3$-statistics (see below).
This unexpected result seems to support an observation \cite{HKO94} that 
the critical exponent $\nu$ of the localization length
is not changed by a magnetic field.
Because of the far reaching consequences for general theories concerning
the MIT, a more careful and comprehensive study is indispensable.

In this letter we present results of numerical investigations which
demonstrate that there is a {\em system size independent crossover} between 
the critical orthogonal and the critical unitary statistics which is 
controlled by the magnitude of an applied Aharonov-Bohm (AB) flux.
Besides the AB-flux model, we investigate two further mechanisms
for breaking time reversal symmetry:
a homogeneous magnetic field and a spatially randomly fluctuating
magnetic flux with zero mean (random flux model).
The same limiting critical unitary level spacing statistics $P_c^u(s)$ is
found in all three cases. Our results also show unambiguously that there 
exists a critical statistics for 3d unitary disordered systems that is
clearly different from the critical orthogonal one. 

The dynamics of non-interacting electrons in a 3d disordered system in the
presence of processes that break the time reversal symmetry can be
investigated by using the Anderson Hamiltonian
\begin{equation}
{\cal H} = \sum_{\bf r} \epsilon_{\bf r} |{\bf r}\rangle\langle{\bf r}|
+ \sum_{\bf r\ne r'} V_{{\bf r,r'}} |{\bf r}\rangle\langle{\bf r'}|.
\end{equation}
The vectors ${\bf r}$ denote the sites of a simple cubic lattice with lattice
constant $a$, and periodic boundary conditions are applied in all directions.
The uncorrelated random energies $\epsilon_{\bf r}$ are distributed with 
constant probability within the interval $-W/2 \le \epsilon_{\bf r} \le W/2$, 
where $W$ denotes the magnitude of the disorder.

The non--diagonal transfer matrix elements between nearest neighbors,
$V_{\bf r,r'}$, contain the symmetry breaking term:
({\it i\/})  for a constant magnetic field $B$,
$V_{\bf r,r'}=V\exp(\mp i2\pi\alpha n)$, if ${\bf r-r'}=\pm
{\bf e}_x$, with ${\bf r}\cdot{\bf e}_y=n$, and $V_{\bf r,r'}=V$ else.
The magnetic field is chosen to be commensurate with the lattice.
({\it ii\/}) for a random magnetic flux
$V_{\bf r,r'}= V\exp(i2\pi\theta_{\bf r,r'})$, where the random
phases $\theta_{\bf r,r'}$ are drawn from an interval
$[-\theta_0/2, \theta_0/2]$ with uniform probability
$p(\theta_{\bf r,r'})=1/\theta_0$.
({\it iii\/}) for the Aharonov-Bohm flux model
$V_{\bf r,r'}= V\exp(\pm i2\pi\phi a/L)$,
${\bf r-r'}=\pm {\bf e}_j$, and $j=x,y,z$,
where the AB-flux $\phi_{\rm AB}=\phi\phi_0$ is given in units of the
flux quantum $\phi_0=h/e$, and $L$ is the linear size of the system.
$V$ is the unit of energy.
\begin{figure}
\begin{center}
\unitlength1.cm
\begin{picture}(8.6,6.6)
\put(0,0){\epsfbox{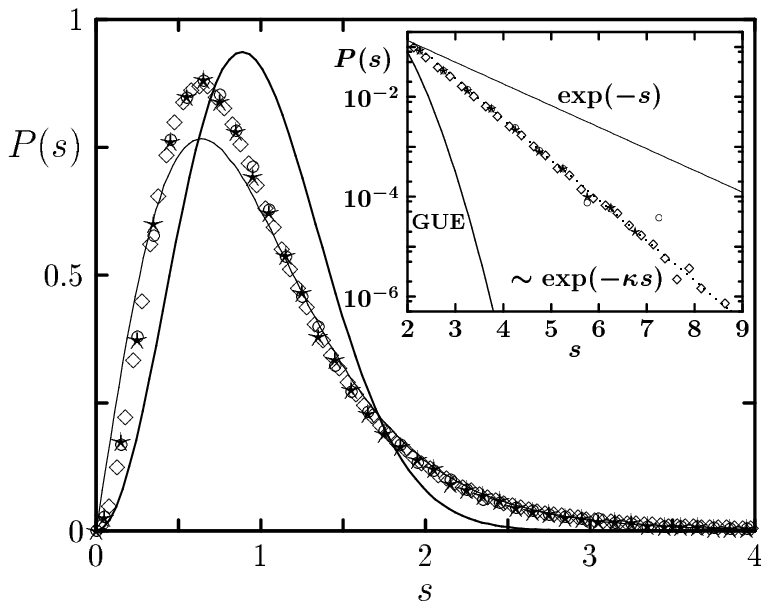}}
\label{fig1}
\end{picture}

\vspace{.5cm}
\parbox{8.5cm}{\small FIG.~1. The critical unitary energy spacing distribution
$P_c^u(s)$ in the presence of a magnetic field ($\alpha=1/5$) for systems of
size $L/a=5$ ($\diamond$), $L/a=10$ ($\star$), and $L/a=20$ ($\circ$).
For comparison, the critical orthogonal $P_c^o(s)$ and the
GUE result are also shown.
The inset shows the large $s$ behavior which can be fitted by an
exponential decay $\exp(-\kappa s)$ with $\kappa=1.87$ (dotted line).}
\end{center}
\end{figure}

Energy eigenvalues were obtained for ensembles of systems containing up to
$32^3$
sites by direct diagonalization (using a Lanczos algorithm for large systems)
or by means of the decimation method \cite{Lee79,BZK95}.
The calculation was restricted to energy intervals about $E/V=0$ 
containing between 3\% and 30\% of the eigenvalues. Within this range we
found no influence on our results. 
The number of realizations depends on the system size. The total number
of eigenvalues amounts to $3\cdot10^{6}$ ($L/a=5$) and $\approx 10^{5}$ for
$L/a=10, 20$.
A proper unfolding procedure has been applied to the energy spectrum
in order to compensate for variations in the density of states.
The normalized spacing distribution $P(s)$ describes the fine-scale spectral 
structure and reflects for $s\to 0$ the level repulsion. 
We also calculated the Dyson-Mehta statistics $\Delta_3$ \cite{Meh91}
which measures the spectral rigidity and describes the spectral structure 
over a range of scales.

In the presence of time reversal symmetry, the critical disorder
at ($E/V=0$) corresponds to $W_c/V\simeq 16.4$ \cite{ZK95a,HoS93b,MacK94}.
It separates localized ($W > W_c$) from metallic states ($W < W_c$).
We find that this critical disorder does not change when an AB-flux is
applied. In contrast, $W_c$ increases with increasing strength of a magnetic 
field \cite{Khm81,Sha84,HKO94}.

We start to present our results with case ({\it i\/}), a 3d-Anderson model
with applied strong constant magnetic field corresponding to
$\alpha=eBa^2/h=1/5$ flux quanta per plaquette.
The critical disorder is $W_c/V=18.1$ \cite{HKO94}.
No dependence of the level spacing distribution on the system
size could be detected within the error bars for $L/a=5, 10, 20$ as
expected for the critical ensemble. Similar results have also been obtained
for magnetic field strengths corresponding to $\alpha=1/4, 1/10$ and $1/20$.

Fig.~1 displays the critical level spacing distribution $P_c^u(s)$ 
in the presence of a constant
magnetic field in comparison with the critical data without magnetic field.
The difference in the heights of the maxima and in the small-$s$ behavior is
clearly visible. This difference is much larger than the small fluctuations
in the numerically obtained distributions.
The small-$s$ behavior is quadratic as expected for the critical unitary case,
in contrast to the linear increase in the critical orthogonal ensemble.
We tried to fit a suggested \cite{AKL94} critical level spacing distribution
$P_{\sc akl}(s)=As^{\beta}\exp(-Bs^{\gamma})$ to our data
with $\gamma=1+1/(\nu d)$,
where $\nu$ and $d$ are the critical exponent of the localization length
and the Euclidean dimension, respectively. While it is possible to fit the
bulk of our data with $\gamma\approx 1.2$
which leads to $\nu\approx 1.66$, the deviation for larger $s$ is obvious.
The large-$s$ behavior, which is shown on a logarithmic scale in the inset of
Fig.~1, is best fitted by an exponential decay $\sim \exp(-\kappa s)$,
with $\kappa\approx 1.9$ having approximately the same value as in the
critical orthogonal case \cite{ZK95b}. A similar exponential decay with
$\kappa \approx 4$ has been reported for the 2d symplectic system
\cite{SZ95}.

The $\Delta_3$-statistics is defined as the least square deviation of the 
integrated density of states, $N(E)$, from a linear behavior, averaged over
an energy range $k$ times the mean level spacing $\Delta$ centered at $E$
\begin{equation}
\Delta_3(k)=\left\langle\min_{A,B} \frac{1}{k}
\int\limits_{-k/2}^{k/2}(N(E+\varepsilon)-A-B\varepsilon)^2\,d\varepsilon
\right\rangle,
\end{equation}
where $\langle\ \rangle$ denotes an ensemble average.
The result shown in Fig.~2 for an energy interval around the 
critical point is again independent of the system sizes studied.
The rigidity of the spectrum is enhanced in the critical unitary as compared
to the critical orthogonal ensemble. The difference between the two
critical curves is significantly larger than the error bars, but smaller
than the difference between GOE and GUE. The influence of the magnetic field
to make the spectrum more rigid in the metallic regime, is weakened at the
critical point due to stronger disorder.
Our results for $P(s)$ and the
$\Delta_3$-statistics clearly indicate the existence of a critical unitary
statistics in the case of an applied strong magnetic field.
\begin{figure}
\begin{center}
\unitlength1.cm
\begin{picture}(8.6,6.8)
\put(0,0){\epsfbox{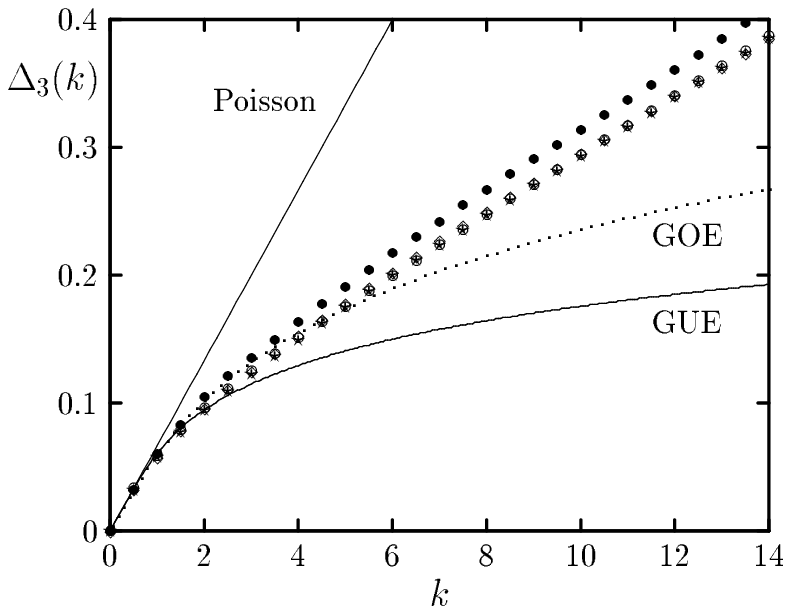}}
\end{picture}

\vspace{.5cm}
\parbox{8.5cm}{\small FIG.~2. The $\Delta_3$-statistics of the critical
unitary ensemble ($W_c/V=18.1, \alpha=0.2$)
for system sizes $L/a=5$ ($\diamond$), $L/a=10$ ($\star$), and $L/a=20$
($\circ$) in comparison with the critical orthogonal case, $L/a=20$
($\bullet$). The known behavior of the insulating (Poisson) and metallic
phases (GOE and GUE) are also shown.}
\end{center}
\end{figure}

In order to investigate whether or not 
a weak magnetic field changes this behavior one must
consider a different model, because due the requirement of
commensurability the system size has to be taken to infinity
as $B\to 0$.
Therefore, we used the random flux model ({\it ii\/}) which allows
to study the effect of weak local magnetic fields with  global magnetic
field being zero. The strength of these local magnetic fields can be tuned
by varying $\theta_0$, the width of the interval from which the random
phases are chosen.
We find the same critical unitary distribution $P_c^u(s)$
as in the strong magnetic field case, provided the condition $L/l_b \gg 1$ is
fulfilled. The magnetic length $l_b$ which serves as a measure for the
strength of the random local magnetic fields can be defined in analogy with 
the usual magnetic length $l_B^2=\hbar/(eB)=a^2/(2\pi\alpha)$.
It is related to the standard deviation, $b$, of the local magnetic fields,
\begin{equation}
l_b^2=\hbar/(e b)=\frac{a^2}{2\pi\left(1/L^3\sum_{\bf r}(\Phi_{\bf r}^p/\phi_0)
^2\right)^{1/2}},
\end{equation}
where the magnetic flux $\Phi_{{\bf r}}^p$ through a plaquette at site
{\bf r}, e.g.
within the $xy$-plane, is calculated from the corresponding random phases
$\Phi_{{\bf r}}^p=\phi_0(\theta_{{\bf r}+{\bf e_x},{\bf r}}+
\theta_{{\bf r}+{\bf e_x}+{\bf e_y},{\bf r}}+
\theta_{{\bf r}+{\bf e_y},{\bf r}+{\bf e_x}+{\bf e_y}}+
\theta_{{\bf r},{\bf r}+{\bf e_y}})$.
The total flux per cross-section perpendicular to each of the three lattice
axis is $\Phi_{\rm T}=\sum_{\bf r}\Phi_{\bf r}^p=0$.

We now turn to the Aharonov-Bohm flux model ({\it iii\/}) for which the absence
of a distinct critical unitary statistics has been asserted \cite{HoS94b}.
Here, we have to distinguish whether a magnetic flux is applied along one,
two or three perpendicular directions. Depending on the magnitude of the flux
and the number of flux directions we observe {\it different scale
independent\/} level statistics for each set of parameters.
To illustrate this unexpected behavior, the change of the maximum $P_{m}(\phi)$
of the critical level spacing distributions is shown
in Fig.~3, where the normalized maximum
$\Gamma_{\rm max}(\phi)=(P_m(\phi)-P_m^o)/(P_m^u-P_m^o)$
is plotted versus the Aharonov-Bohm flux $\phi$.
We find a smooth crossover starting from
the maximum of the critical orthogonal $P_m^o$ to the maximum of the
critical unitary ensemble $P_m^u$ when the flux is
increased along two or three directions.
Hence, in the limit of large flux there exists the same critical unitary
level statistics $P_c^u(s)$ known from the random flux model and from the
strong magnetic field case also for the 2- and 3-component AB-flux model.
For one flux line in a single direction, however, this critical unitary
distribution is not reached.
Increasing the magnetic flux beyond $\phi=0.25$ decreases
$P_m(\phi)$ because of a $\phi_0/2$-periodicity in the flux: at $\phi=0.5$
all phases in $V_{\bf r,r'}$ are equal to unity and the corresponding
Hamiltonian is real.
This behavior is known as ``false T-breaking'' \cite{RB86}.
In all cases, within the error bars, the results are independent of the
system sizes investigated.
We find a similar crossover also for the $\Delta_3$-statistics.
\begin{figure}
\begin{center}
\unitlength1.cm
\begin{picture}(8.6,6.8)
\put(0,0){\epsfbox{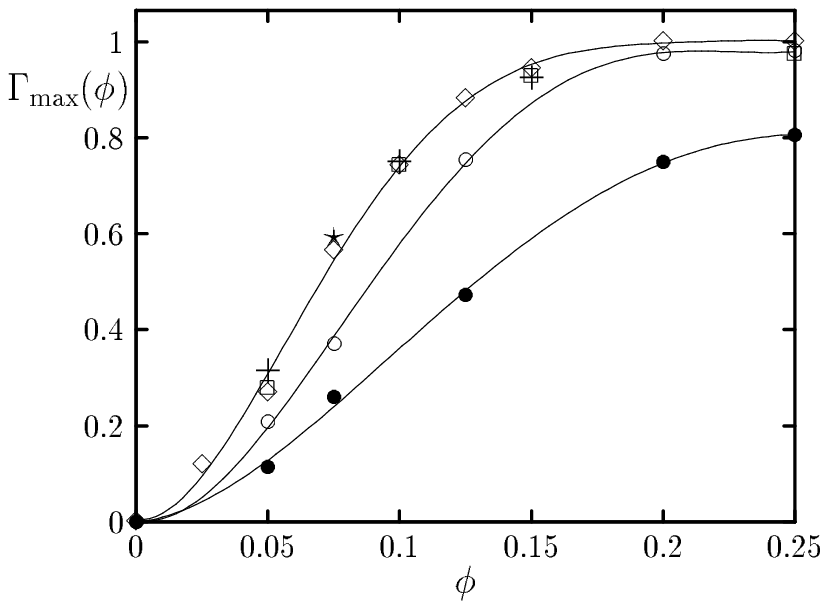}}
\end{picture}

\vspace{.5cm}
\parbox{8.5cm}{\small FIG.~3. The normalized maximum,
$\Gamma_{\rm max}(\phi)$,
of the critical level spacing distribution versus the Aharonov-Bohm flux
$\phi$ showing the system size independent crossover from critical
orthogonal to the critical unitary ensemble. For AB-flux applied in three
perpendicular directions data are shown for $L/a=5$
($\Diamond$), $L/a=10$ ($+$), $L/a=20$ ($\Box$), and $L/a=32$
({\large$\star$}).
For AB-flux in two ($\circ$) and one ($\bullet$) direction only
the results for $L/a=5$ are shown.}
\end{center}
\end{figure}

The small-$s$ behavior of the critical $P(s,\phi)$ for various values
of the AB-flux applied in all three directions is shown in Fig.~4. 
While for small
$\phi$ the overall deviation of the level spacing distribution from the
critical orthogonal curve is hardly visible, the small-$s$ dependence is
clearly quadratic. This holds for all fluxes considered, although
the range of $s$ values showing the $s^2$-behavior decreases with decreasing
flux.  Only for $\phi=0$ the linear relation, $P(s)\sim s$,
characteristic of the orthogonal ensemble, is obtained.
\begin{figure}
\begin{center}
\unitlength1.cm
\begin{picture}(8.6,6.8)
\put(0,0){\epsfbox{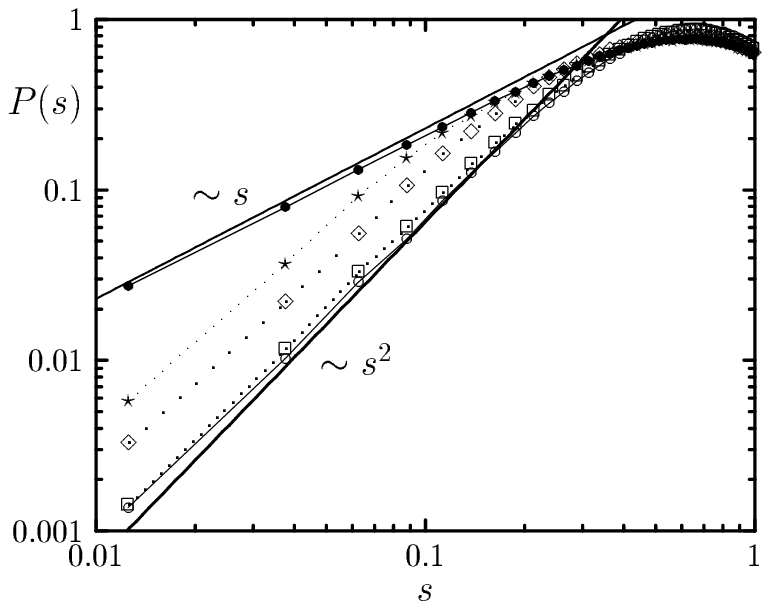}}
\end{picture}

\vspace{.5cm}
\parbox{8.5cm}{\small FIG.~4. Double logarithmic plot of the small-$s$
behavior of $P(s)$ for various Aharonov-Bohm fluxes $\phi=0.0$
($\bullet$), $\phi=0.025$ ($\star$), $\phi=0.05$ ($\diamond$),
$\phi=0.125$ ($\Box$), $\phi=0.25$ ($\circ$).}
\end{center}
\end{figure}

In the metallic phase, only the basic symmetry determines the 
characteristics of the spectral fluctuations even though the mechanisms to
break the time reversal symmetry differ.
In a magnetic field each closed electron-orbit not lying in 
a plane parallel to the magnetic field direction picks up a magnetic flux,
which is in contrast to the smaller set of those closed trajectories
with winding number larger than zero, e.g. trajectories that circle round the
single flux line in the AB-flux model. The latter situation was shown to be 
the reason for a flux controlled GOE $\to$ GUE crossover in Aharonov-Bohm
chaotic billiards \cite{BR86}. Nevertheless, this crossover becomes abrupt in 
the semiclassical limit $L\to \infty$. A similar discontinuous transition is
known from RMT \cite{Meh91} for infinite order matrices and also from 
numerical studies on metallic samples \cite{DM91} where the conductance
diverges with $L$.

Our results for the MIT show that the critical unitary statistics 
depend on how the time reversal symmetry is broken. This is due to the 
{\em finite} critical conductance in combination with the system 
size independent Aharonov-Bohm flux $\phi_{\rm AB}=\phi h/e$. Hence, 
the crossover becomes scale invariant at the metal-insulator transition point. 
This is different in the magnetic field case, because here the magnetic flux 
$\phi_{\alpha}=BL^2$ increases with system size.

In conclusion, we have investigated the energy level statistics at the
critical point of a 3d Anderson model when time reversal symmetry is
broken by either a constant magnetic field, a spatially fluctuating magnetic
flux, or an Aharanov-Bohm flux.
A critical unitary level spacing distribution has been found that  
is distinct from the critical orthogonal one.
We further showed that there exists a scale invariant AB-flux
controlled crossover regime so that the critical
level statistics depend on how the time reversal symmetry is broken.

\bibliographystyle{prsty}

\end{multicols}
\end{document}